\documentclass[slac_one]{revtex4}
\usepackage{graphicx}
\usepackage{fancyhdr}
\usepackage{amsmath}
\newcommand{\vl}{V_{\rm L}}
\newcommand{\vr}{V_{\rm R}}
\newcommand{\gl}{g_{\rm L}}
\newcommand{\gr}{g_{\rm R}}

\begin{document}

\title{Recent studies of top quark properties and decays at hadron colliders} 

%

\author{V. Chiochia}
\affiliation{University of Zurich, Physik-Institut, 8057 Z\"urich, Switzerland}
%

\begin{abstract}
The top quark is the heaviest known elementary particle.  Observed for the first time in 1995 at the Tevatron by the CDF and D0 experiments, it has become object of several studies aimed at fully characterize its properties and decays. Precise determinations of top quark characteristics verify the internal consistency of the standard model and are sensitive to new physics phenomena. With the advent of the large top quark production rates generated at the LHC, top quark studies have reached unprecedented statistical precision. This review summarizes the recent measurements of top quark properties and studies of its decays performed at the LHC and Tevatron.  
\end{abstract}

\maketitle

\thispagestyle{fancy}

\section{Introduction}

The top (or $t$) quark is the heaviest and the most recently discovered quark in the standard model (SM) of elementary particles. Observed for the first time in 1995 at the Tevatron collider by the CDF and D0 experiments~\cite{Abe:1995hr,Abachi:1995iq}, it has a mass of $173.2 \pm 1.0$ GeV~\cite{TeVcomb}, about 40 times heavier than its partner, the bottom (or beauty, $b$) quark and more than twice heavier than the W boson. It decays almost exclusively to a W boson and a bottom quark with an expected lifetime of about $10^{-25}$ s. The lifetime is too short to allow the formation of bound states with other quarks, such as hadrons,  and the top quark properties at production time are directly transferred to the its decay products.

The top quark can be produced at two hadron colliders: the Tevatron at Fermilab, colliding protons with antiprotons at center-of-mass energies up to $\sqrt{s}=1.96$~TeV, and at the Large Hadron Collider (LHC), colliding protons at $\sqrt{s}=7$ and 8 TeV. Between 2001 and 2011 the Tevatron experiments collected an integrated luminosity of approximately 10.5~fb$^{-1}$ during the so-called Run II. The LHC delivered the first proton collisions in 2009 and the 2010-11 run at $\sqrt{s}=7$~TeV provided data samples corresponding to an integrated luminosity of about 5.2~fb$^{-1}$. A run at $\sqrt{s}=8$~TeV  started in April 2012 and is expected to deliver an integrated luminosity of about 20~fb$^{-1}$ by the end of 2012. After a 18 months shutdown period, the LHC will restart operation at the center-of-mass energy of 13~TeV.

At hadron colliders the top quark can be produced in pairs through strong interactions, or singly via weak interactions. At the Tevatron with $\sqrt{s}=1.96$~TeV, about 85\% of the total production rate is expected to be from the annihilation of the valence quarks and antiquarks, while the remaining 15\% is from gluon-gluon fusion. The expected  cross section for pair production computed at next-to-leading order accuracy (NLO) of quantum chromodynamics (QCD) is $\sigma_\mathrm{t\bar{t}}=6.74 ^{+0.52}_{-0.80}$~pb for a top mass $m_t=173$~GeV~\cite{Kidonakis:2011ca}. Production of single top or antitop quarks proceeds through the exchange of a W boson in the s- or t-channel, or in the $Wt$ associated production channel. At the Tevatron the production cross section for single top (or antitop) is $3.4\pm 0.2$~pb. At the LHC top pair production is largely dominated by gluon-gluon fusion (80\% at $\sqrt{s}=7$~TeV) while the rest is from $q\bar{q}$ annihilation. The cross section for pair production is more than 20 times larger than the corresponding Tevatron value, $\sigma_\mathrm{t\bar{t}}=160 ^{+22}_{-23}$~pb at $\sqrt{s}=7$~TeV~\cite{Ahrens:2011px}, making the LHC a real top factory. The single top cross section is about 40\% of the pair production at $\sqrt{s}=7$~TeV and is not equal for quarks and antiquarks due to the proton parton density functions. The single top cross section is about a factor 2 larger than the corresponding antitop cross sections in the t- and s-channels, while the $Wt$ cross section is the same for $t$ and $\bar{t}$.

In this review the recent results on top quark properties and decays obtained at the Tevatron and the LHC are summarised. The results are obtained from the analysis of Tevatron Run II data and LHC data collected in 2011. The measurements are performed with the CDF, D0, CMS and ATLAS experiments, described in Refs.~\cite{CDFdetector,D0detector,CMSdetector,ATLASdetector}. 
The article is structured as follows: in Section~\ref{sec:top-mass} the measurements of the top quark mass and $\mathrm{t\bar{t}}$ mass difference are summarised; the determination of the W boson helicity fractions in top decays and associated constraints on the $Wtb$ vertex coupling are reported in Section~\ref{sec:Whel}; the measurements of $\mathrm{t\bar{t}}$ spin correlations and charge asymmetries are summarised in Section~\ref{sec:spincorr} and \ref{sec:chargeasym}, respectively, while Section~\ref{sec:FCNC} describes the current status of searches for rare top decays involving flavour-changing neutral currents. 
%
%
\section{Top quark mass and $\mathrm{t\bar{t}}$ mass difference\label{sec:top-mass}}

The top quark mass is a free parameter of the SM. It determines the Yukawa coupling of the quark to the Higgs boson and contributes to several electroweak precision observables via radiative corrections. Precise measurements of the top quark and W boson masses can constrain the Higgs boson mass and represent an important internal consistency test of the SM. 
The ATLAS and CMS experiments have recently announced the observation of a new boson with mass of about 125-126~GeV~\cite{:2012gu,:2012gk}. Further studies and larger datasets are required to fully establish the nature of the new particle. If confirmed as the Higgs boson of the SM, the precise determination of its mass and of the top quark mass, along with the strong coupling constant, will constraint the extrapolation of electroweak vacuum stability up to the Planck scale~\cite{Alekhin:2012py}.
The main methodology used to determine the top quark mass at hadron colliders consists in measuring the invariant mass of the decay products of the top event candidates and deducing the mass, $m_\mathrm{top}$, using sophisticated analysis methods. The most precise measurements of this type use the $t\bar{t}\rightarrow$ lepton + jets channel, where one of the W bosons from the $\mathrm{t\bar{t}}$ final state decays into a charged lepton and a neutrino and the other into a pair of quarks. The CMS and ATLAS collaborations have recently performed measurements using this decay channel as well as dileptonic final states, where both W bosons decay leptonically.  

\begin{figure}[htbp]
\begin{center}
\includegraphics[width=5.3cm]{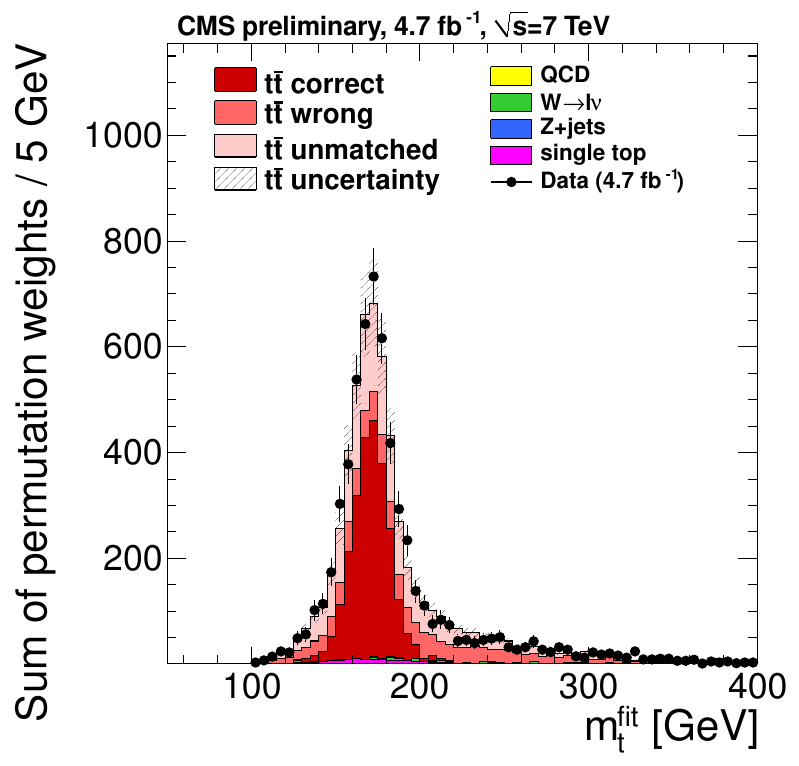}
\includegraphics[width=7.6cm]{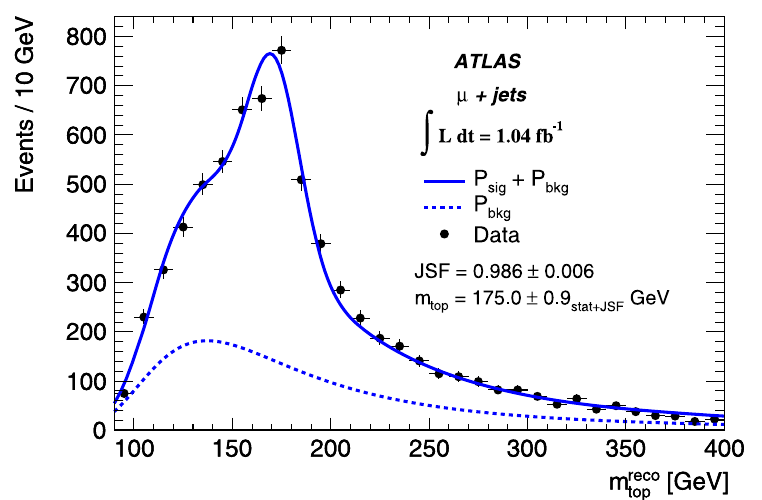}\hspace{-1mm}
\includegraphics[width=4.75cm]{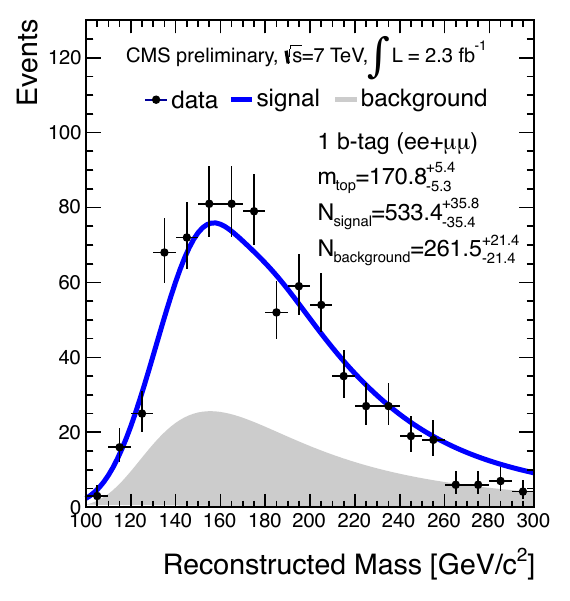}
\caption{Measurements of the top quark mass in final states with one or more leptons. {\it Left:} Distribution of the top quark mass from a kinematic fit measured by CMS in the $\mu$+jets channel. The stacked histogram shows the Monte Carlo expectation for a mass of 172.5~GeV. The shaded area represents the normalization uncertainty from the CMS cross section measurement. {\it Middle:} Mass distribution fitted to the ATLAS data for the $\mu$+jets channel. The dashed lines denote the background and the full lines the sum of the signal and background templates. {\it Right:} Reconstructed mass distribution from the CMS dilepton analysis, for leptons pairs of the same flavor. The solid line shows the result of the fit.\label{fig:TOP-11-015}}
\end{center}
\end{figure}

In the CMS lepton plus jets analysis~\cite{TOP-11-015}, based on 4.7~fb$^{-1}$, events with one muon with $p_T>30$~GeV and $|\eta|<2.1$ and at least four jets with $p_t>30$~GeV and $|\eta|<2.4$ are selected. At least two b-tagged jets are requested. The jet energy scale (JES) and top quark mass are extracted simultaneously from an event based 2-D likelihood function, also known as {\it ideogram}. A kinematic fit is employed in order to test the kinematic compatibility of an event with the $\mathrm{t\bar{t}}$ hypothesis and improve the resolution on the measured quantities by exploiting the knowledge of the decay process. A cut on the fit probability, $P_\mathrm{fit}>0.2$, is imposed to enhance the fraction of correct permutations. The top mass distribution from the kinematic fit is shown in Figure~\ref{fig:TOP-11-015} and the measured value is reported in Table~\ref{tab:top-mass}. The most relevant sources of systematic uncertainties are the b-jet JES, the factorization scale and the pile-up effects.

The ATLAS lepton plus jets measurement~\cite{ATLAS:2012aj} is based on events with one lepton candidate (muon or electron) and at least four jets. The processed integrated luminosity is 1.04~fb$^{-1}$. The lepton must have $E_T>25$~GeV for electrons and $p_T>20$~GeV for muons. The missing transverse energy is required to be $E_T^\mathrm{miss}>20~(35)$~GeV for the muon (electron) channel. The jets are required to have $p_T>25$~GeV and $|\eta|<2.5$ and at least one of the four jets must be tagged as b-jet. The top quark mass is extracted using the template method, where templates are fitted to functions that interpolate between different input values of the physics observable, fixing all other parameters of the functions. In the final step a likelihood fit to the observed data distribution is used to obtain the value for the physics observable that best describes the data (see Figure~\ref{fig:TOP-11-015}). Two implementations of the template method are performed: a one-dimensional analysis based on the observable $R_{32}$ and a two-dimensional analysis determining simultaneously $m_\mathrm{top}$ and the JES, similarly to the CMS case. The top quark mass extracted from the combination of the two lepton channels is reported in Table~\ref{tab:top-mass}. A combined result from 1d- and 2d-analyses does not improve the precision of the measured top quark mass from the 2d-analysis. The most relevant systematic uncertainties are from the relative b-jet to light jet energy scale, the modeling of initial and final state QCD radiation, and the light quark jet energy scale.

The CMS collaboration has also performed a measurement of the top mass in dileptonic decays using an integrated luminosity of 2.3~fb$^{-1}$~\cite{TOP-11-016}. Of all $\mathrm{t\bar{t}}$ decays the dilepton channel has the smallest branching fraction and it is expected to be the least contaminated by background processes. Events are selected with two isolated, oppositely charged leptons (electrons or muons) with $p_T>20$~GeV and $|\eta|<2.4$, and at least two jets with $p_T>30$~GeV and $|\eta|<2.4$. The missing transverse energy must be $E_T^\mathrm{miss}>30$~GeV and at least one b-tagged jet is required. The top quark mass is measured using the KINb method, in which the kinematic equations describing the $\mathrm{t\bar{t}}$ system are solved many times per event for each lepton-jet combination. The combination with the largest number of solutions is chosen and the mass value, $m_\mathrm{KINb}$, is estimated by a Gaussian fit around the most probable value. An two-component, unbinned likelihood fit of the $m_\mathrm{KINb}$ distribution is performed to estimate $m_\mathrm{top}$ (see Figure~\ref{fig:TOP-11-015}). The measurement is the most precise estimate in the dilepton channel to date and is reported in Table~\ref{tab:top-mass}. The dominant source of systematic uncertainty on the top mass is the JES.

The possible decay channels of $\mathrm{t\bar{t}}$ pairs include the case in which both W bosons decay to quark-antiquark pairs. This mode yields a six quark final state, two of which are b quarks from the top-quark decays. Additional quarks or gluons may be created by initial or final state radiation from the interacting partons. Measurements using fully hadronic final states were recently reported by the ATLAS and CDF collaborations. The ATLAS analysis, based on 2~fb$^{-1}$, selects events with five jets of $p_T>55$~GeV and a sixth jet of $p_T>30$~GeV~\cite{CONF-2012-030}. To suppress backgrounds from multijet events and collinear b-quark pair production from gluon splitting two of the reconstructed jets with $|\eta|<2.5$ are required to be b-tagged and separated by a three-dimensional angular aperture $\Delta R>1.2$. The top quark mass, reported in Table~\ref{tab:top-mass}, is obtained by fitting templates to the three-jet mass combination of the selected jet triplets. The templates are obtained by combining simulated signal samples for five top masses in the range 160-190 GeV with a background contribution obtained from data by mixing jets from different events. The dominant systematic uncertainties on the top mass are the inclusive and b-jet JES, the background modelling and the contribution of initial- or final-state radiation. The CDF analysis in hadronic top decays is based on 5.8~fb$^{-1}$~\cite{CDF-10456}. Events with six to eight jets with $E_T>15$~GeV and minimum angular distance $\Delta R>0.5$ are processed with a neural network (NN) using 13 input kinematic variables. Events are further selected requiring the NN output above a given threshold and one to three b-tagged jets. The top mass (see Table~\ref{tab:top-mass}) and JES are simultaneously extracted from a 2-D template method, using templates with top quark masses between 160 and 185~GeV. The most relevant sources of systematic uncertainties on the mass measurement are the background modelling, Monte Carlo modelling of the hard interaction and residual JES uncertainties.
%
%
\begin{table}[htdp]
\caption{Summary of recent top quark mass measurements performed at hadron colliders}
\begin{center}
\begin{tabular}{cccc}
\hline 
Method & Experiment & $m_\mathrm{top}$ (GeV) & Ref. \\ \hline
$\mu+$jets & CMS   & $172.6 \pm 0.6\mathrm{(stat.+JES)} \pm 1.2\mathrm{(syst.)}$ & \cite{TOP-11-015} \\
$e,\mu+$jets (1D) & ATLAS & $174.4 \pm 0.9\mathrm{(stat.)} \pm 2.5\mathrm{(syst.)}$ & \cite{ATLAS:2012aj} \\
$e,\mu+$jets (2D) & ATLAS & $174.5 \pm 0.6\mathrm{(stat.)} \pm 2.3\mathrm{(syst.)}$ & \cite{ATLAS:2012aj} \\
Dilepton          & CMS   & $173.3 \pm 1.2\mathrm{(stat.)} ^{+2.5}_{-2.6} \mathrm{(syst.)}$ & \cite{TOP-11-016} \\
Hadronic          & ATLAS & $174.9 \pm 2.1\mathrm{(stat.)} \pm 3.8\mathrm{(syst.)}$ & \cite{CONF-2012-030} \\
Hadronic          & CDF   & $172.5 \pm 1.4\mathrm{(stat.)} \pm 1.4\mathrm{(syst.)}$ & \cite{CDF-10456} \\
\hline
\end{tabular}
\end{center}
\label{tab:top-mass}
\end{table}%

\begin{figure}[htbp]
\begin{center}
\includegraphics[width=9cm]{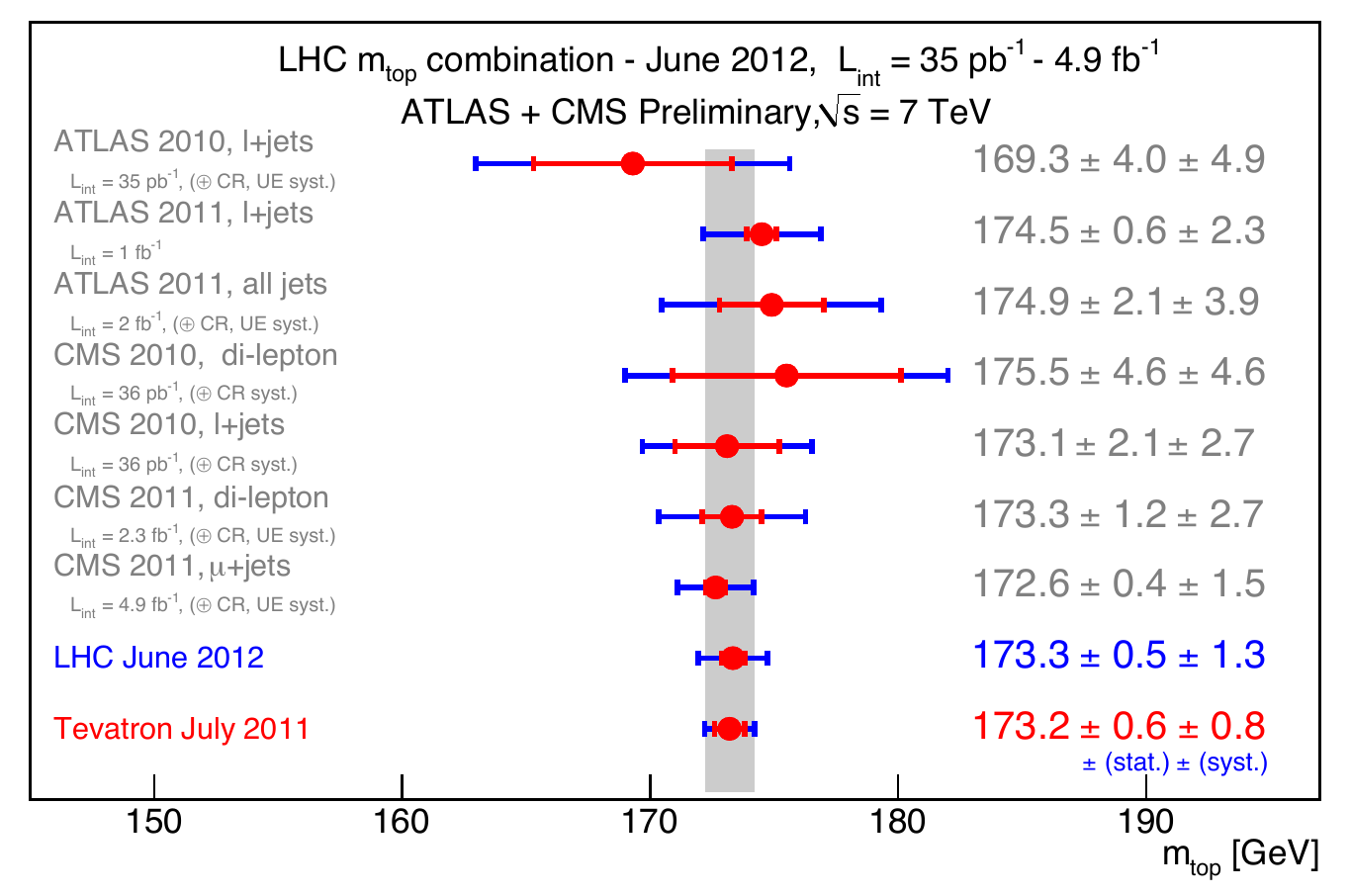}
\caption{Input measurements and result of the LHC top quark mass combination, compared to the corresponding Tevatron result.}
\label{fig:LHCcomb}
\end{center}
\end{figure}

To make best use of the single top quark mass measurements the results are combined. At the Tevatron, measurements by the CDF and D0 collaborations based on Run I and Run II data, corresponding to integrated luminosites up to 5.8~fb$^{-1}$, have been combined yielding $m_\mathrm{top}=173.2 \pm 0.6\mathrm{(stat.)} \pm 0.8\mathrm{(syst.)}$~GeV~\cite{TeVcomb}. The LHC measurements have been recently combined using the Best Linear Unbiased Estimator (BLUE) method~\cite{LHCcomb}. The combination includes 2010 and 2011 measurements in lepton+jets, dilepton and all hadronic decays, based on integrated luminosities  between 35~pb$^{-1}$ and 4.9~fb$^{-1}$. The resulting combination, taking into account statistical and systematic uncertainties and their correlations, yields $m_\mathrm{top}=173.3 \pm 0.5\mathrm{(stat.)} \pm 1.3\mathrm{(syst.)}$~GeV~\cite{TeVcomb}, as shown in Figure~\ref{fig:LHCcomb}. The value is consistent with the Tevatron combination and the total uncertainty is currently dominated by the systematics uncertainties due to the jet calibration, the signal modelling and the underlying event tune. Although the top quark mass is now known to sub-percent accuracy, future measurements are expected to increase the precision even further. With the larger datasets collected during the 2012 LHC run the event selections can be limited to the phase-space regions where the detector effects are better understood and data-driven techniques will better constrain the ranges of parameter variations adopted in the systematic uncertainties determination.

Due to CPT conservation the mass of a quark must be equal to that of its antiparticle. The top quark is the only case in which a free quark can be observed, for it decays before hadronizing, constituting an excellent test bench for CPT conservation. The mass difference between the top and its antiquark, $\Delta m_t = m_t-m_{\bar{t}}$, was measured by the D0 and CDF experiments, showing no significant deviation from zero~\cite{Aaltonen:2011wr}. The values measured by CDF and D0 are $\Delta m_t=-3.3\pm 1.7$~GeV and $0.84 \pm 1.87$~GeV, respectively. The CMS experiment has recently measured the mass difference  with unprecedented precision~\cite{Chatrchyan:2012ub}. The analysis, based on an integrated luminosity of about 5~fb$^{-1}$, selects events with one W boson candidate decaying into quark-antiquark pairs and the other decaying leptonically, where the lepton is a muon or a neutrino. The data sample is divided into negative and positive lepton charges. For each category, the ideogram method is used to measure the mass of the top or anti top quark and the difference between the masses in the two categories of lepton charge is taken as the mass difference. The measured mass difference is $\Delta m_t = -0.44 \pm 0.46\mathrm{(stat.)} \pm 0.46 \mathrm{(syst.)}$~GeV.  The largest systematic uncertainties are currently given by the pile-up effects, the method calibration and the response difference for b anti-b jets. The measurement is in agreement with the the assumption of CPT invariance and is more precise by at least a factor three than any of the previous measurements.

%
%
\section{W boson helicity in top decays and $\mathrm{Wtb}$ coupling\label{sec:Whel}}

In the SM the helicity of the W boson produced in a $t \rightarrow Wb$ decay may be longitudinal, left- or right-handed. The helicity fractions are defined as the partial rate of a given helicity state divided by the total decay rate: $F_{L,R,0} = \Gamma_{L,R,0}$, where $F_L$, $F_R$ and $F_0$ are the left-handed, right-handed and longitudinal helicity fractions, respectively. The top quark is produced with negligible polarisation at hadron colliders and decays via the $V-A$ charged-current interaction, which strongly suppresses right-handed $W^+$ bosons or left-handed $W^-$ bosons. The expression of the helicity fractions depends on the W boson, top and bottom quark masses. For fixed values of the masses, $M_W=80.399$~GeV, $m_t=173.3$~GeV and $m_b=5$~GeV, the fractions predicted at next-to-next-to-leading-order (NNLO) QCD are known with sub-percent precision: $F_0=0.687\pm 0.005$, $F_L=-0.311\pm 0.005$ and $F_R = (1.7\pm 0.1) \times 10^{-3}$~\cite{Czarnecki:2010gb}. The small and reliable SM prediction makes $F_R$ a particularly sensitive probe for new physics. Deviations from the expectations could in fact indicate non-SM contributions to the $Wtb$ vertex, as we will discuss in Section~\ref{sec:Wtbcoupling}.

Measurements of the W helicity fractions have been performed both at Tevatron and at the LHC. The helicity is measured through the study of angular distributions of top decay products in different decay channels. The helicity angle $\theta^*$ is defined as the angle between the charged lepton momentum in the W rest frame and the W momentum in the top rest frame. The CDF and D0 experiments have performed measurements of of $F_0$ and $F_R$ in the lepton+jets and dilepton channels. The measurements are of two types: (i) a model-independent approach (also called "2D") where $F_0$ and $F_R$ are determined simultaneously and (ii) a model-dependent approach here $F_0$ ($F_R$) is fixed to its SM value and $F_R$ ($F_0$) is measured (called "1D"). In both cases the fractions are assumed to satisfy the condition $F_R+R_0+F_L=1$. The single measurements from both approaches have been recently combined using the BLUE method~\cite{Aaltonen:2012rz}. The values obtained from the combination procedure are summarized in Table~\ref{tab:Whelicity}.
%
%
\begin{table}[htdp]
\caption{Measurements of the W helicity fractions in top decays at hadron colliders. For each result the first uncertainty is statistical and the second systematic. The SM prediction is shown in the last row.}
\begin{center}
\begin{tabular}{cccccc} \hline
Method & Experiment & $F_0$ & $F_R$ & $F_L$ & Ref. \\ \hline
2D & D0+CDF & $0.722 \pm 0.062 \pm 0.052$ & $-0.033 \pm 0.034 \pm 0.031$ & - & \cite{Aaltonen:2012rz}\\
1D & D0+CDF & $0.682 \pm 0.035 \pm 0.046$ & $-0.015 \pm 0.018 \pm 0.031$ & - & \cite{Aaltonen:2012rz}\\
3D & CMS &    $0.567 \pm 0.074 \pm 0.047$ & $-0.040 \pm 0.035 \pm 0.044$ & $0.393 \pm 0.045 \pm 0.029$ & \cite{TOP-11-020}\\
2D & CMS &    $0.643 \pm 0.034 \pm 0.050$ & 0 & $0.357 \pm 0.034 \pm 0.050$& \cite{TOP-11-020}\\
Combined & ATLAS & $0.67 \pm 0.03 \pm 0.06$ & $0.01 \pm 0.01 \pm 0.04$ & $0.32 \pm 0.02 \pm  0.03$ & \cite{Aad:2012ky} \\
\hline
\multicolumn{2}{c}{SM prediction (NNLO)} & $0.687\pm 0.005$ & $(1.7\pm 0.1) \times 10^{-3}$ & $-0.311\pm 0.005$ & \cite{Czarnecki:2010gb} \\ \hline
\end{tabular}
\end{center}
\label{tab:Whelicity}
\end{table}%

The CMS experiment has measured the W helicity fractions using a dataset corresponding to an integrated luminosity of 2.2~fb$^{-1}$~\cite{TOP-11-020}. Events are selected with one W candidate decaying hadronically and the other W decaying leptonically, into a muon and a neutrino. Muons with $p_T>25$~GeV and at least four jets with $p_T>30$~GeV are requested. The W+jets background is suppressed by requiring at least one of the jets to be b-tagged. The $\mathrm{t\bar{t}}$ event is reconstructed via a constrained kinematic fit. Only events with at least four jets and one lepton are used, with constrains optimised to improve the $\cos(\theta^*)$ resolution. Two types of fits are performed: (i) a "3D fit" where the fractions $F_0$, $F_L$ and the $\mathrm{t\bar{t}}$ component normalisation, $F_{\mathrm{t\bar{t}}}$,  are treated as free parameters and $F_R=1-F_0-F_L$, and (ii) a "2D fit" leaving $F_0$ and $F_{\mathrm{t\bar{t}}}$ as free parameters and setting $F_R=0$, $F_L=1-F_0$. The results of the two methods are reported in Table~\ref{tab:Whelicity}.

The measurement of W helicity fraction was performed also by the ATLAS experiment, with a data sample corresponding to an integrated luminosity of 1.04~fb$^{-1}$~\cite{Aad:2012ky}. Events with at least one isolated lepton (electron or muon) are selected with $E_T>25$~GeV ($p_T>20$~GeV) for electrons (muons). In the single lepton channels four jets are requested with at least one of them being tagged as b-jet. In the dilepton channels ($ee$, $e\mu$ and $\mu \mu$) at least two jets are required, with at least one of them being b-tagged. The helicity fractions were extracted with two methods using events reconstructed with a kinematic kit: a  fit of the $\cos(\theta^*)$ distribution using different W helicity templates from simulation, and from a measurement of the angular asymmetries using an unfolded $\cos(\theta^*)$ spectrum corrected for background contributions. The single results were combined using the BLUE method, taking into account the statistical correlations between analyses and the correlations of systematic uncertainties (see Table~\ref{tab:Whelicity}). All experimental results are in agreement with NNLO QCD predictions with the ATLAS result being the most precise estimate from a single experiment.

\subsection{Constraints to the anomalous $Wtb$ couplings\label{sec:Wtbcoupling}}

Any deviation of the helicity fractions from the SM prediction could be caused by new physics contributing to the $Wtb$ vertex. New physics can be parameterised in terms of an effective Lagrangian above the electroweak symmetry breaking scale:
\begin{equation}
\mathcal{L}_{Wtb} = - \frac{g}{\sqrt 2} \bar b \, \gamma^{\mu} \left(
\vl P_{\rm L} + \vr P_{\rm R}
\right) t\; W_\mu^- - \frac{g}{\sqrt 2} \bar b \, \frac{i \sigma^{\mu
\nu} q_\nu}{M_W}
\left( \gl P_{\rm L} + \gr P_{\rm R} \right) t\; W_\mu^- + \mathrm{h.c.} \,,
\label{ec:lagr}
\end{equation}
where $P_{\rm L}$ ($P_{\rm R}$) is the left-handed (right-handed) 
chirality operator and
\begin{align}
& \vl = V_{tb} + C_{\phi q}^{(3,3+3)} \frac{v^2}{\Lambda^2} \,,
&&\vr = \frac{1}{2} C_{\phi \phi}^{33*} \frac{v^2}{\Lambda^2} \,,
& \gl = \sqrt 2 C_{dW}^{33*} \frac{v^2}{\Lambda^2} \,,
&& \gr = \sqrt 2 C_{uW}^{33} \frac{v^2}{\Lambda^2} \,.
\label{eq:anom_couplings}
\end{align}
The parameter $\Lambda$ is the scale of new physics while $C_{\phi q}^{(3,3+3)}$, $C_{\phi \phi}^{33*}$, $C_{dW}^{33*}$ and $C_{uW}^{33}$ are the effective operator coefficients. The anomalous couplings $\vr$, $\gl$, $\gr$, generated by dimension-six operators, are equal to zero in the SM, while the coupling
$V_{tb}$ receives a correction from the operator $O_{\phi q}^{(3,3+3)}$~\cite{Aad:2012ky}.
\begin{figure}[htbp]
\begin{center}
\includegraphics[width=7cm]{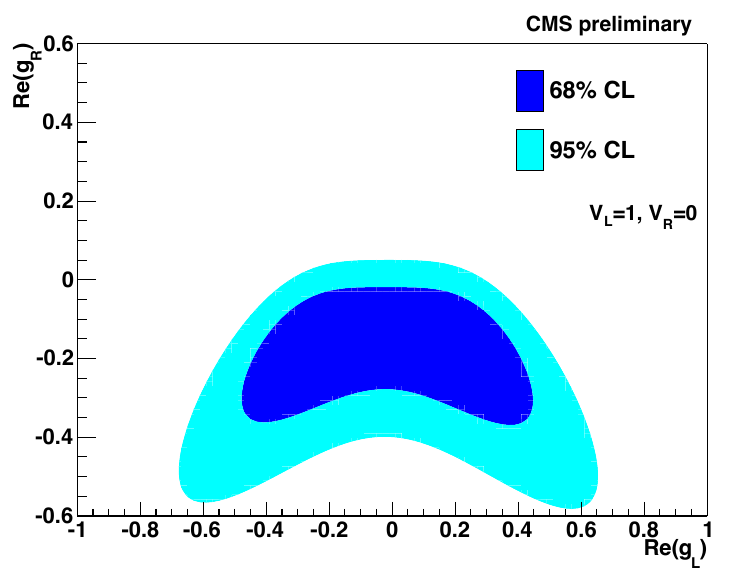}
\includegraphics[width=7.4cm]{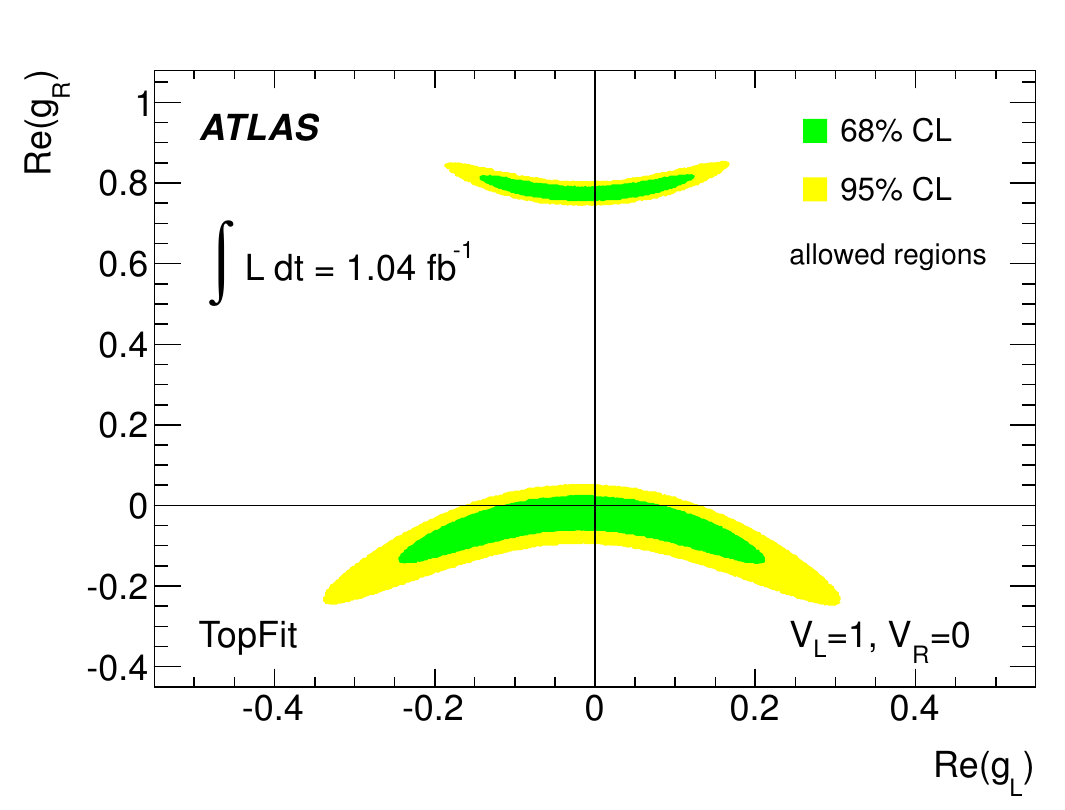}
\caption{Allowed regions at 68 and 95 \% CL in the Re$(\gl)$, Re$(\gr)$ plane from the CMS (left) and ATLAS (right) measurements of the W helicity fractions in top decays.}
\label{fig:Wtbcoupling}
\end{center}
\end{figure}

The CMS experiment has set limits on the anomalous $Wtb$ couplings in two specific scenarios~\cite{TOP-11-020}: (i) assuming $\vl=1, \vr=\gr=0$ and leaving $\mathrm{Re}(\gr)$ as free parameter and (ii) leaving free the possibility of observing a non-zero value of $F_R$. The first approach yields the best fit at Re$(g_R)=-0.070 \pm 0.053 \mathrm{(stat.)} ^{+0.073}_{-0.081} \mathrm{(syst.)}$, corresponding to a coefficient $\mathrm{Re}(C_{uW}^{33})/\Lambda^2 = -0813 \pm 0.615 \mathrm{(stat.)} ^{+0.847} _{-0.951} \mathrm{(syst.)}$~TeV$^{-2}$. The allowed regions at 68 and 95\% confidence level (CL) in the Re$(\gl)$, Re$(\gr)$ plane from the second approach are shown in Figure~\ref{fig:Wtbcoupling} together with the corresponding ATLAS result. The one-dimensional 95\% CL intervals extracted by ATLAS are $-0.20<\mathrm{Re}(V_R)<0.23$, $-0.14<\mathrm{Re}(\gl)<0.11$ and $-0.08<\mathrm{Re}(\gr)<0.04$. It should be noted that the region of Re$(\gr)\simeq 0.8$ is disfavoured by single-top cross section measurements at Tevatron and LHC and is not shown in the CMS figure. The LHC results are consistent with the $(V-A)$ structure of the $Wtb$ vertex and improve on the limits previously obtained at Tevatron~\cite{:2012iwa}.
%
%
\section{Top antitop spin correlations\label{sec:spincorr}}

Thanks to the small decay width, about one order of magnitude smaller than the scale of strong interactions ($\Lambda_{\mathrm{QCD}}\simeq 0.1$~GeV), the spin of the top quark is directly transferred to decay products and can be measured from their angular distributions. At hadron colliders  top quark pairs are expected to be produced with spins significantly correlated. The observation of spin decorrelation could indicate that the spin orientation was modified by a non-SM decay process, such as a top decaying into a scalar charged Higgs boson and a b-quark ($t \rightarrow H^+ b$). Measurements of spin correlations were first performed by the CDF and D0 experiments, however none of these measurements had sufficient sensitivity to discriminate between the SM and the uncorrelated spins hypothesis. More recently, the D0 experiment has performed a measurement based on an integrated luminosity of 5.3~fb$^{-1}$, providing the first evidence of spin correlations in $\mathrm{t\bar{t}}$ events with a significance exceeding three standard deviations~\cite{Abazov:2011gi}. The degree of correlation, $A$, can be defined as the fractional difference between the number of events where the top and anti top quark spin orientations are aligned and those where the top quark spins have opposite alignment. The value $A=+1$ ($-1$) corresponds to fully parallel (antiparallel) spins. Next-to-leading-order (NLO) QCD predicts $A=0.78^{+0.03}_{-0.04}$ for $p\bar{p}$ collisions at $\sqrt{s}=1.96$~TeV~\cite{Bernreuther:2004jv}. The D0 measurement is based on a matrix-element approach using lepton+jets events and has been combined with the measurement in the dilepton channel. One isolated electron or muon with $p_T>20$~GeV and four jets with $p_T>20$~GeV are requested, where the leading jet has $p_T>40$~GeV. The ratio of events with correlated spins to the total number of events, $f_\mathrm{meas}$, is extracted from a binned maximum-likelihood fit to the discriminant distribution using simulated templates of $\mathrm{t\bar{t}}$ with and without spin correlations. The combination of the two channels yields $f_\mathrm{meas}=0.85\pm 0.2 \mathrm{(stat.+syst.)}$, which excludes the region $f<0.344$ at 95\% CL. 
\begin{figure}[htbp]
\begin{center}
\includegraphics[width=6cm]{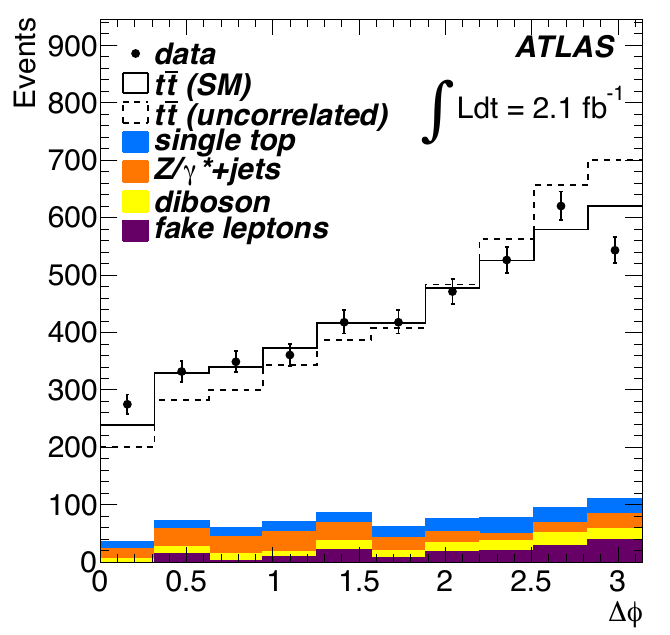}
\includegraphics[width=6.4cm]{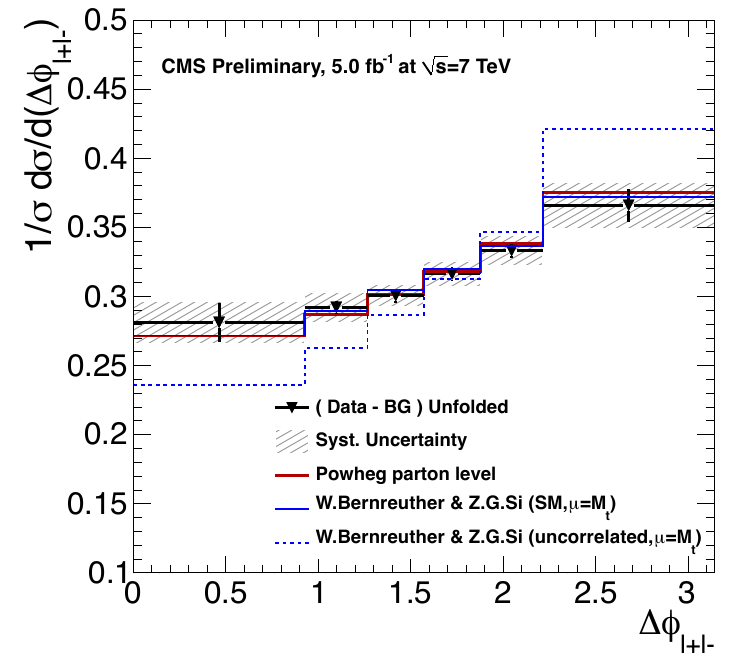}
\caption{Reconstructed charged lepton $\Delta \phi$ distribution in dileptonic $\mathrm{t\bar{t}}$ decays from ATLAS (left) and CMS (right). The CMS measurement is background-subtracted and unfolded to parton-level.}
\label{gif:spincorr}
\end{center}
\end{figure}

Further studies of $\mathrm{t\bar{t}}$ spin correlations were recently reported by the ATLAS and CMS experiments. At the LHC top pair production occurs mostly through the gluon-gluon channel. At low invariant masses, $m_\mathrm{t\bar{t}}\le400$~GeV, the production process is dominated by fusion of gluon pairs of equal helicity, resulting in $\mathrm{t\bar{t}}$ pairs with aligned orientations of the top and anti-top spins (left-left or right-right)~\cite{Mahlon:2010gw}. When both W bosons from the top pair decay leptonically they produce charged leptons with correlated azimuthal angle in the laboratory frame, a quantity well measured by LHC detectors. The ATLAS measurement, based on an integrated luminosity of 2.1~fb$^{-1}$, selects events with exactly two oppositely charged lepton candidates ($ee$, $\mu\mu$, $e\mu$)~\cite{ATLAS:2012ao}. The electron (muon) candidates are selected with $p_T>25 (20)$~GeV  and dilepton invariant mass above 15 GeV~\cite{Mahlon:2010gw}. An additional cut on the missing transverse energy is imposed on events with lepton pairs of the same flavor to suppress background from Z+jets and W+jets events. The spin correlation is extracted from a binned log-likelihood fit of the dilepton azimuthal angle difference, $\Delta \phi$. The signal template is constituted by a linear superimposition of the SM $\mathrm{t\bar{t}}$ simulation with coefficient $f^{SM}$, and from the uncorrelated $\mathrm{t\bar{t}}$ simulation with coefficient $(1-f^{SM})$. Thus, $f^{SM}=0$ corresponds to uncorrelated spins. Figure~\ref{gif:spincorr} shows the reconstructed $\Delta \phi$ distribution compared with the SM expectation and uncorrelated $\mathrm{t\bar{t}}$ simulation.  The measured value from the fit is $f^{SM} = 1.30 \pm 0.14 \mathrm{(stat.)} ^{+0.27}_{-0.22} \mathrm{(syst.)}$, where the largest systematic uncertainties are associated to the JES, jet resolution and efficiency, and fake leptons. The measurement can be translated into a determination of the correlation degree, $A$, by multiplying for a factor obtained from NLO QCD and dependent on the spin basis. The result in the helicity basis is reported in Table~\ref{tab:spincorr} together with the SM prediction at NLO accuracy.
\begin{table}[htdp]
\caption{LHC measurements of the spin correlation coefficient in the helicity basis. The SM prediction is shown in the last row.}
\begin{center}
\begin{tabular}{ccc}
Experiment & $A_\mathrm{helicity}$ & Ref. \\ \hline
ATLAS      & $0.40 \pm 0.04 \mathrm{(stat.)} ^{+0.08}_{-0.07} \mathrm{(syst.)}$ & \cite{ATLAS:2012ao} \\ 
CMS        & $0.24 \pm 0.02 \mathrm{(stat.)} \pm \mathrm{0.08(syst.)}$          & \cite{TOP-12-004}\\
\hline
SM prediction (NLO) & 0.31 & \cite{Bernreuther:2010ny} \\
\hline
\end{tabular}
\end{center}
\label{tab:spincorr}
\end{table}%

A similar analysis was recently performed by the CMS experiment, based on an integrated luminosity of 5 fb$^{-1}$~\cite{TOP-12-004}. Candidate events are selected with two isolated charged leptons (electrons or muons) with $p_T>20$~GeV and two or more jets, where at least one is identified as originating from a b-quark. The events are also required to have a dilepton invariant mass above 20 GeV for the three channels and a missing transverse energy above 40 GeV for the $ee$ and $\mu\mu$ channels only. A binned likelihood fit of the reconstructed $\Delta \phi$ distribution is used to extract the spin correlation coefficient. All the events are fitted together in a single  fit, but different templates are used according to the decay channel. The background-subtracted and unfolded differential cross-section as function of $\Delta \phi$ is shown in Figure~\ref{gif:spincorr}. The result in the helicity basis is in good agreement with the SM expectation and is reported in Table~\ref{tab:spincorr}. The largest systematic uncertainties are associated to the background normalisation and MC statistics. The CMS experiment also investigated the spin correlations through asymmetry distributions, inclusively and at high $m_\mathrm{t\bar{t}}$ at reconstruction level, and inclusively unfolded to parton level. These results are also in agreement with the SM predictions.
%
%
\section{Charge asymmetries\label{sec:chargeasym}}

At its lowest order QCD predicts charge symmetric distributions for top quark pairs produced at hadron colliders. At NLO the interference between the Born diagram and the box diagram, as well as between initial- and final-state gluon emission, correlates the flight directions of the top quarks and antiquarks to the directions of motion of the initial quarks and antiquarks, respectively. The $\mathrm{t\bar{t}}$ production in gluon-gluon fusion is however, charge symmetric. At Tevatron top pairs are predominantly produced by quark-antiquark annihilation, therefore top quarks are emitted preferentially along the direction of motion of the incoming protons and the top antiquarks along the direction of the antiprotons. This leads to a small forward-backward asymmetry of $(6 \pm 1)\%$ in the $\mathrm{t\bar{t}}$ rest frame~\cite{Almeida:2008ug,Antunano:2007da,Bowen:2005ap}. The signed difference between the rapidities of the top and anti-top, $\Delta y = y_t - y_{\bar{t}}$, reflects the asymmetry in $\mathrm{t\bar{t}}$ production and one can define the integrated charge asymmetry as $A_{\mathrm{fb}} = (N_f-N_b) / (N_f+N_b)$, where $N_f$ ($N_b$) is the number of events with a positive (negative) $\Delta y$. Both CDF and D0 experiments have performed measurements of $A_{\mathrm{fb}}$. Using samples of about 5~fb$^{-1}$, CDF measured parton level asymmetries of $A_{\mathrm{fb}} = (15.8 \pm 7.4)\%$~\cite{Aaltonen:2011kc} in the lepton+jets channel and $A_{\mathrm{fb}} = (42 \pm 16)\%$~\cite{CDFNote10436} in the dilepton channel. Combining the two CDF results yields $A_{\mathrm{fb}} = (20.1 \pm 6.7)\%$~\cite{CDFNote10584}. A measurement performed by the D0 experiment in the lepton+jets channel and based on an integrated luminosity of 5.4~fb$^{-1}$ gives $A_{\mathrm{fb}} = (19.6 \pm 6.5)\%$, in good agreement with CDF~\cite{Abazov:2011rq}. More recently, CDF performed a new measurement based on 8.7~fb$^{-1}$\cite{CDFNote10807}. The measured inclusive asymmetry is $[16.2 \pm 4.1\mathrm{(stat.)} \pm 2.2\mathrm{(syst.)}]\%$ while the NLO QCD prediction from {\tt POWHEG} Monte Carlo is $A=6.6\%$. CDF has also measured the asymmetry differentially as function of the rapidity difference $\Delta y$ and top anti-top invariant mass, $m_\mathrm{t\bar{t}}$. In both cases, the asymmetry dependence on the measured variable is consistent with a linear increase, but the measured slopes are about 2.4$\sigma$ larger than the NLO QCD predictions. 

The unexpectedly large forward-backward asymmetry observed at the Tevatron can be further tested at the LHC. In the latter case, $\mathrm{t\bar{t}}$ production via $q\bar{q}$ or $qg$ scattering is expected to be small, and the dominant process is $gg \rightarrow t \bar{t}$. The symmetric initial state does not generate forward-backward asymmetries and $A_{\mathrm{fb}}$ is not a useful observable any longer. However, the rapidity distribution of top and anti-top pairs differs, with top quarks emitted more frequently towards larger rapidities. The effect is due to the different momenta of the incoming $q\bar{q}$ partons as the valence quarks carry, on average, a larger fraction of the proton momentum than the anti-quark from the proton sea. The anti-top quark distribution is, therefore, more peaked towards central rapidity values. The charge asymmetry can be measured from the difference of the absolute values of pseudorapidities (or rapidities), $\Delta |\eta| = |\eta_t| - |\eta_{\bar{t}}|$. Taking $N^{+(-)}$ as the number of events with positive (negative) values of $\Delta |\eta|$ the charge asymmetry is defined as \begin{equation}A_C=(N^+-N^-)/(N^++N^-).\label{eq:asym}\end{equation} The SM prediction for the LHC case as obtained from the {\tt MC@NLO} Monte Carlo generator is $A_C = (0.6 \pm 0.2)\%$, where the uncertainty is estimated from  the variation of the renormalization and factorization scales and using different sets of PDF. The prediction in Ref.~\cite{Kuhn:2011ri}, taking into account electroweak effects and setting the denominator in Eq.~\ref{eq:asym} to leading order accuracy, yields $(1.15 \pm 0.06)\%$~\cite{ATLAS:2012an}.

\begin{table}[htdp]
\caption{Summary of the $\mathrm{t\bar{t}}$ charge asymmetry measurements at the LHC.}
\begin{center}
\begin{tabular}{ccccc} 
Experiment & Method & Int. luminosity (fb$^{-1}$) & $A_C$ (\%) & Ref. \\ \hline
ATLAS & $e,\mu$+jets & 1.0 & $-1.9 \pm 2.8\mathrm{(stat.)} \pm 2.4\mathrm{(syst.)}$ & \cite{ATLAS:2012an} \\
ATLAS & dilepton      & 4.7 & $ 5.7 \pm 2.4\mathrm{(stat.)} \pm 1.5\mathrm{(syst.)}$ & \cite{ATLAS-2012-057} \\ \hline
ATLAS & combined     & 1--4.7 & $ 2.9 \pm 1.8\mathrm{(stat.)} \pm 1.4\mathrm{(syst.)}$ & \cite{ATLAS-2012-057} \\ \hline
CMS & $e,\mu$+jets & 4.7 & $0.4 \pm 1.0\mathrm{(stat.)} \pm 1.2\mathrm{(syst.)}$ & \cite{TOP-11-030} \\
\end{tabular}
\end{center}
\label{tab:LHCA_C}
\end{table}%

Measurements of $A_C$ have been performed by both ATLAS and CMS experiments using lepton+jets and dilepton decays. In the ATLAS measurement~\cite{ATLAS:2012an}, based on about 1~fb$^{-1}$, events are selected with a single isolated electron (muon) with $p_T>25~(20)$~GeV, at least four jets with $p_T>25$~GeV and missing transverse energy $E_T^{\mathrm{miss}}>20$~GeV. In the 'tagged' selection at least one of the jets is required to be b-tagged. After background subtraction, a Bayesian unfolding procedure is performed to correct for acceptance and detector effects. The result is in agreement with the SM expectation and is reported in Table~\ref{tab:LHCA_C}. The largest systematic uncertainties are associated to the modeling of the $\mathrm{t\bar{t}}$ signal and gluon radiation as well as the multi-jet background for the electron channel. More recently ATLAS has reported a measurement in the dilepton channel based on 4.7~fb$^{-1}$~\cite{ATLAS-2012-057}. Events are selected with two oppositely-charged isolated muons (electrons) with $p_T>20~(25)$~GeV, a missing transverse energy $E_T^\mathrm{miss}>60$~GeV and two jets with $p_T>25$~GeV. Lepton pairs of different flavor are also included. A method based on leading-order matrix elements is adopted to fully reconstruct the $\mathrm{t\bar{t}}$ four momenta, by fixing the top quark and W boson masses to the world average values. The asymmetry in each channel is measured after background subtraction and after correction for acceptance and detector effects.  The combination of the three channels ($ee$, $\mu\mu$ and $e\mu$) based on the BLUE method is reported in Table~\ref{tab:LHCA_C} and is found to be consistent with the SM expectation. The dominant systematic uncertainties on the single measurements are the multi-jet background modeling as well as the electron efficiency and resolution. The ATLAS experiment has also measured a lepton-based asymmetry, $A_C^{ll}$, that does not require the reconstruction of the full $\mathrm{t\bar{t}}$ system. In this case, the measured discriminating variable is $\Delta |\eta| = |\eta_{l^+}| - \eta|_{l^-}|$, representing the difference of the absolute values of positively and negatively charged lepton pseudorapidities. The result is $ A_C^{ll} = [2.3 \pm 1.2\mathrm{(stat.)} \pm 0.8\mathrm{(syst.)}]\%$, in good agreement with the $\tt MC@NLO$ Monte Carlo prediction of $(0.4\pm 0.1)\%$.

The CMS experiment has measured the charge asymmetry is lepton+jets decays with a data sample corresponding to an integrated luminosity of about 1~fb$^{-1}$~\cite{Chatrchyan:2011hk}. The result has been recently updated with 4.7~fb$^{-1}$~\cite{TOP-11-030}. Events are selected with an isolated electron or muon with $p_T>25$~GeV and 17~GeV, respectively, together with at least three jets, each with $p_T>30$~GeV. The reconstructed top quark and antiquark four-vectors are used to obtain the inclusive and differential distributions of $\Delta |y|$ as function of the rapidity, transverse momentum and invariant mass of the $\mathrm{t\bar{t}}$ system. The corresponding asymmetry $A_C$ is also measured differentially from Eq.~\ref{eq:asym}. The inclusive measurement is reported in Table~\ref{tab:LHCA_C} where the dominant systematic uncertainties are associated to the unfolding technique and to the lepton identification and efficiency. Both the inclusive and differential distributions are consistent with the SM predictions and no hints for contributions from physics beyond the SM are found.
\begin{figure}[htbp]
\begin{center}
\includegraphics[width=7cm]{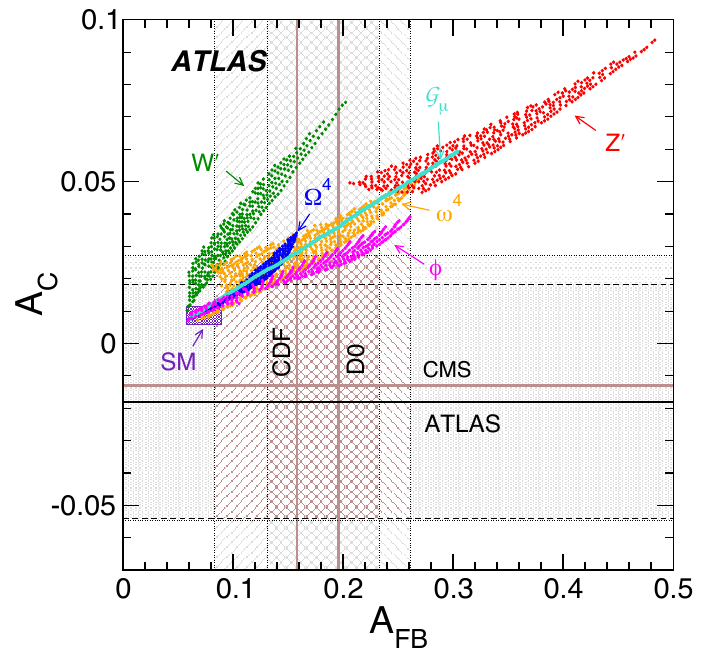}
\caption{Forward-backward asymmetries from the Tevatron (vertical bands) and charge asymmetries  from the LHC (horizontal bands), compared to the SM prediction and various models including new physics contributions (colored areas).}
\label{fig:BSMasym}
\end{center}
\end{figure}

The forward-backward asymmetries measured at the Tevatron and the charge asymmetries from the LHC have been compared to the predictions of a number of simple models beyond the SM~\cite{ATLAS:2012an}, as shown in Figure~\ref{fig:BSMasym}. The colored areas represent the ranges of predicted values for $A_{FB}$ and $A_C$ for the new physics models, computed using the tree-level SM amplitude and adding the contributions of new particles. The colored box is the SM prediction. The horizontal bands correspond to the ATLAS and CMS measurements from Refs.~\cite{ATLAS:2012an} and \cite{Chatrchyan:2011hk}, respectively, while the vertical bands represent the Tevatron measurements from Refs.~\cite{Aaltonen:2011kc,Abazov:2011rq}. The LHC charge asymmetry measurements disfavor models with a new flavor-changing $Z'$ boson and are in tension with models introducing a $W'$ boson with right-handed couplings, while for the other new physics models the asymmetries measured at the Tevatron are consistent with the LHC within experimental uncertainties. The same study performed in the region of high invariant mass $m_\mathrm{t\bar{t}}>450$~GeV shows that the predictions of the six new physics models are in tension with the CDF and ATLAS high-mass measurements considered together.
%
%
\section{Searches for flavor-changing neutral currents in top decays\label{sec:FCNC}}

Several extensions of the SM predict that the branching fractions of the top quark decaying to a  Z boson and a lighter quark, $t \rightarrow Zq$, where $q$ is a $u$ or $c$ quark, can be substantially enhanced~\cite{AguilarSaavedra:2004wm}. In the SM this decay occurs at one loop level but is strongly suppressed by the GIM mechanism. The SM expectation for the branching ratio is Br$(t\rightarrow Zc)\simeq 1 \times 10^{-14}$, while the decay  $t \rightarrow Zu$ is further suppressed by a factor $|V_{ub}/V_{cb}|^2\simeq 7.9 \times 10^{-3}$, yielding Br$(t\rightarrow Zu)\simeq 8 \times 10^{-17}$. The branching ratio in the new physics models is typically many orders of magnitude larger than the SM value, and can be as high as $2\times 10^{-4}$ in certain R-parity violating SUSY models~\cite{AguilarSaavedra:2004wm}. Searches for flavour-changing neutral current decay $t \rightarrow Zc$ performed at CDF~\cite{Aaltonen:2008ac} and D0~\cite{Abazov:2011qf} result in the upper limits Br$(t\rightarrow Zc)\leq 3.7\%$ and $\leq 3.2\%$, respectively, at 95\% CL. The larger top production cross section at the LHC allows to substantially improve the Tevatron bounds. 

The CMS and ATLAS experiments have recently performed searches for these rare top decays. The ATLAS measurement, based on 2.1~fb$^{-1}$, searches for events with one top (or antitop) decaying through the $t \rightarrow Wb$ channel while the other decays in the suppressed $Zq$ mode~\cite{Aad:2012ij}. Only leptonic decays of the Z and W bosons are considered ($Z\rightarrow ee,~\mu\mu,~\tau\tau$ and $W\rightarrow e\nu,~\mu\nu,~\tau\nu$), leading to a final state with three isolated leptons, at least two jets and missing transverse energy from the semi-leptonic W boson decay. The events are divided into two categories: (i) a sample with three leptons identified from the combination of inner detector, calorimeter and muon spectrometer, and (ii) a sample in which one of the leptons is based on the inner detector information only while the other two lepton candidates are as in (i). In the former case the leading lepton is required to have $p_T>20$~GeV, and the two sub-leading leptons are required to have $p_T>20$~GeV. Furthermore, two jets with $p_T>25$~GeV are required and in the sample (ii) at least one of the two jets is required to be b-tagged. Finally, the missing transverse energy is required to be $E_T^\mathrm{miss}>20$~GeV. The selected event samples are required to be kinematically consistent with the $t\bar{t} \rightarrow WbZq$ decay through a $\chi^2$ minimisation technique and further cuts are imposed on the reconstructed W and Z boson candidates masses. The number of events after all selection cuts is consistent with the background from SM processes and fake leptons and no evidence of the $Zq$ decay  is found. The expected and observed upper limits extracted with the CL$_\mathrm{s}$ method are reported in Table~\ref{tab:FCNC}.
\begin{table}[htdp]
\caption{Summary of the expected and observed 95\% CL upper limits on the FCNC top quark decay $t \rightarrow Zq$ branching ratio at hadron colliders.}
\begin{center}
\begin{tabular}{cccc}
Experiment & Observed & Expected & Ref. \\ \hline
CDF   & $3.7 \times 10^{-2}$ & $5.0 \times 10^{-2}$& \cite{Aaltonen:2008ac} \\
D0    & $3.2 \times 10^{-2}$ & $3.8 \times 10^{-2}$& \cite{Abazov:2011qf}\\
ATLAS & $7.3 \times 10^{-3}$ & $9.3 \times 10^{-3}$ & \cite{Aad:2012ij} \\
CMS   & $3.4 \times 10^{-3}$ & $3.4 \times 10^{-3}$ & \cite{TOP-11-028} \\
\end{tabular}
\end{center}
\label{tab:FCNC}
\end{table}%

The CMS measurement is based on an integrated luminosity of 4.6~fb$^{-1}$~\cite{TOP-11-028}. Similarly to the ATLAS case, the search is performed by selecting a sample of $t\bar{t} \rightarrow WbZq$ candidate decays, where the W and Z bosons decay to electrons or muons. All three leptons in the final state are required to be isolated and have $p_T>20$~GeV. The invariant mass of and $ee$ or $\mu\mu$ pair is required to be consistent with the Z boson mass. At least two jets with $p_T>30$~GeV are requested, separated by $\Delta R>0.4$ from the lepton candidates. The resulting sample is then divided into two categories: (i) a sample requiring a minimum value of the variable $HT_s$, defined as the scalar sum of the lepton transverse momenta, jet transverse energies and missing transverse energy, and (ii) requiring one of the two jets to be b-tagged. The latter selection is the most sensitive and hence taken as reference result. The number of observed events after all selection cuts is consistent with the expectation from SM background processes and no evidence of the $Zq$ decay is found also in this case. The expected and observed upper limits computed with the CL$_\mathrm{s}$ technique are given in Table~\ref{tab:FCNC} and represent the most stringent bound from a single experiment to date. Furthermore, this result improves on the Tevatron upper limits by approximately an order of magnitude.

%
%
\section{Summary and outlook}

Thanks to the rapid increase of the LHC datasets and unprecedented production rates the field of top quark physics is living a stimulating and intense season with several new precise measurements being achieved. The recent increase of the LHC center-of-mass energy from 7 to 8~TeV has further enhanced the cross sections by 20\% to 50\% depending on the production process. The top quark mass, a fundamental free parameter in the SM, has been determined with sub-percent precision and is the best known quark mass to date. The most precise single determinations have systematic uncertainties of about 1~GeV and statistical uncertainties below 1~GeV. Thus, improved analysis techniques will be required to further reduce the systematic uncertainties. The mass difference between top and anti top quarks is in agreement with hypothesis of CPT conservation and has reached a precision relative to the top quark mass of about 0.4\%.  If the new boson recently observed by LHC experiments \cite{:2012gu,:2012gk} is confirmed as the Higgs boson of the SM, precise measurements of its mass, along with the W boson and top quark masses, will provide stringent consistency tests for the SM and a better understanding of the electroweak vacuum stability at large energy scales.

The $(V-A)$ structure of top decays to W bosons has been probed by measuring the W helicity fractions. LHC experiments have reached sufficient sensitivity to determine all three helicity fractions with precisions up to 10\%. The statistical and systematic uncertainties in this case are roughly of the same order. The measurements are in agreement with SM expectations but still about one order of magnitude less accurate that theoretical predictions. The results are also translated into exclusion limits on the anomalous couplings in the $Wtb$ vertex. 

Thanks to the short lifetime the spin correlation in top antitop decays can be studied from the angular correlations of the decay leptons. Spin correlation were first observed at more than 3$\sigma$ level by the D0 experiment. The LHC experiments have now measured non zero correlation coefficients with an accuracy of about 20-30\%, excluding the zero correlation hypothesis at $3-4.5\sigma$ confidence level.

The study of $t\bar{t}$ forward-backward asymmetries, $A_{FB}$, at the Tevatron has revealed an intriguing enhancement with respect to SM prediction, at the level of 2.4$\sigma$. Due to the different production process dominating the LHC top pair cross section the observable $A_{FB}$ is replaced by alternative approaches, such as the measurement of charge asymmetries. The results from CMS and ATLAS experiment are both in agreement with NLO QCD predictions. The combination of Tevatron and LHC measurement set complementary constraints to models proposed in an attempt to explain the Tevatron data. For instance, models introducing a new flavor-changing $Z'$ boson are disfavoured by LHC measurements.

Searches for rare top decays involving flavor-changing neutral currents have received an immediate boost from the large LHC samples. Although no evidence of the rare decays $t\rightarrow Zq$ is observed the upper limits on these decays have been recently improved by approximately an order of magnitude and are now of the order of 10$^{-3}$. Further improvements of about a factor of two can be expected from the larger dataset collected during the 2012 LHC run.

The LHC will provide the largest top event samples ever collected. With a total integrated luminosity of about 20~fb$^{-1}$ expected by the end of 2012 the LHC experiments will have collected about 40 times more top quark pairs and 60 times more single top events than Tevatron experiments at 10~fb$^{-1}$. The unprecedented datasets will disclose new opportunities for precision measurements and searches for new phenomena.

%
%

\end{document}